\documentclass[useAMS,usenatbib]{mn2e}
\usepackage{amsmath, amssymb, graphicx, hyperref,color}

\newcommand{\lya}{Ly$\alpha$\hspace{1mm}}
\newcommand{\apj}{ApJ}
\newcommand{\aap}{A\&A}
\newcommand{\apjl}{ApJL}
\newcommand{\mnras}{MNRAS}

\newcommand{\pasa}{PASA}

\newcommand{\hs}{\hspace{1mm}}

\def\gsim{~\rlap{$>$}{\lower 1.0ex\hbox{$\sim$}}}

\def\lsim{~\rlap{$<$}{\lower 1.0ex\hbox{$\sim$}}}

\title[Fermi Accelerated Ly$\alpha$]{Blue Wings and Bumps via Fermi-like Acceleration of Ly$\alpha$ Photons across Shocks}
\author[Chung et al.]{Andrew S. Chung$^{1}$\thanks{E-mail:
andy@mpa-garching.mpg.de}, Mark Dijkstra$^{2}$\thanks{E-mail:
mark.dijkstra@astro.uio.no}, Benedetta Ciardi$^{1}$\thanks{E-mail:
ciardi@mpa-garching.mpg.de}, Max Gronke${^2}$\thanks{E-mail:
max.gronke@astro.uio.no}\\
\\$^{1}$Max Planck Institut f\"{u}r Astrophysik, Karl-Schwarzschild-Str. 1, 85741 Garching, Germany
\\$^{2}$Institute of Theoretical Astrophysics, University of Olso, Postboks 1029, 0315 Oslo, Norway}
\begin{document}
\date{\today}

\pagerange{\pageref{firstpage}--\pageref{lastpage}} \pubyear{2013}
\maketitle
\label{firstpage}
\begin{abstract}
We explore the impact of Fermi-like acceleration of Lyman-alpha (Ly$\alpha$) photons across shock fronts on the observed \lya spectral line shape.
We first confirm the result of Neufeld \& McKee (1988) that this mechanism gives rise to extended blue wings which may have been observed in some radio galaxies.
Our Monte-Carlo radiative transfer calculations further show that in a minor modification of the shell-model, in which we add an additional static shell of hydrogen, this process can naturally explain the small blue bumps observed in a subset of Ly$\alpha$ emitting galaxies, which have been difficult to explain with conventional shell-models. Blue bumps can be produced with an additional column density of static hydrogen as small as $N^{\rm static}_{\rm HI} \ll  N^{\rm shell}_{\rm HI}$, and typically occur at roughly the outflow velocity of the shell.
In our model the spectra of so-called `blue-bump objects' might reflect an evolutionary stage in which the outflows regulating the escape of Ly$\alpha$ photons are still engulfed within a static interstellar medium.

\end{abstract}
\section{Introduction}
\label{introduction}

Observations indicate that the escape of Ly$\alpha$ photons is facilitated enormously by the presence of outflowing interstellar gas \citep[e.g.][]{Kunth98}. Scattering off these outflows effectively Doppler shifts Ly$\alpha$ photons out of resonance into the red wing of the absorption line profile. In models of this process the outflow is often represented with a geometrically thin shell of gas \citep[e.g.][]{Verhamme2006,Schaerer2011}. In spite of its simplicity, this so-called `shell-model' has been very successful at reproducing observed spectra line shapes \citep{Verhamme2008,Hashimoto15}. 

However, the shell-model has recently been shown to have difficulties, especially in reproducing the strength of `blue bumps' in a subset of observed spectra \citep{Kulas12,Chonis13,Hashimoto15}. \citet{Adams} studied what may be viewed as an extreme example of blueshifted Ly$\alpha$ emission in which the Ly$\alpha$ line of spatially extended Ly$\alpha$ emission around a radio galaxy as a whole is blueshifted relative to the observed 21-cm absorption, with emission extending beyond 1500 km s$^{-1}$ into the blue wing of the line profile. 

Blueshifted Ly$\alpha$ radiation arises naturally when the photons scatter through optically thick inflowing gas \citep{Zheng02,D06a,D06b}. Indeed, \citet{Adams} show that their data can be reproduced if more than $10^{12} M_{\odot}$ of cold inflowing gas is present. This large mass of cold neutral gas inside a massive dark matter halo ($M \gsim 10^{13} M_{\odot}$) though, is at odds with expectations from theory, which predicts that the gas should be predominantly accreted in the hot-mode (e.g. \citealt{Keres05}; Fig.~7 of \citealt{DB06}).

An alternative process which gives rise to blueshifted Ly$\alpha$ emission is described by Neufeld \& McKee (1988; henceforth NM88), who show that scattering across a shock front can give rise to such blueshift. NM88 presented analytical calculations, and therefore were forced to adopt simplifying assumptions, namely that the shock-crossing acceleration process is modelled by \lya photons bouncing between two partially transparent, frequency-preserving mirrors. In reality, however, the frequency of a \lya photon is not preserved at each `reflection'. Instead, the frequency of the photon diffuses through frequency-space as the photon is scattered by the material on either side of the shock front, in turn changing the optical depth of the gas to the photon. In addition, they assume that photons cross the shock front isotropically. Finally, the analysis presented by NM88 was restricted to a simplified geometry of two adjacent semi-infinite slabs.

In spite of an increasing prevalence of Ly$\alpha$ radiative transfer Monte-Carlo codes \citep[][]{Zheng02,Cantalupo2005,D06a,Tasitsiomi2006,Verhamme2006,Laursen2009,Yajima2012}, they have so far not been used to investigate this Fermi-like acceleration mechanism (while acknowledging the inaccuracy of doing so, we henceforth refer to this as `Fermi acceleration' for brevity).
The goals of this paper are two-fold: ({\it i}) numerically study Fermi acceleration of Ly$\alpha$ photons across a shock front without the simplifying assumptions that were required in the analytical treatment, and ({\it ii}) investigate whether this mechanism can help to reproduce observed blue bumps in spectra that have been difficult to explain with conventional shell-models. The outline of this paper is as follows: \S~\ref{mechanism} describes the basic acceleration mechanism, \S~\ref{simulations} details the simulations which we use to investigate the physical mechanism, \S~\ref{simResults} presents the output of our numerical experiments, \S~\ref{applications} discusses the applications of our findings, and in \S~\ref{conclusions} we summarise the conclusions.

\section{Blueshifting Mechanism}\label{mechanism}

\lya photons are resonantly scattered by HI, making the \lya radiative transfer complex. The propagation of \lya photons is affected by both bulk gas motion and, in detail, microscopic motion of individual hydrogen atoms. The result is that \lya photons undergo a random-walk like motion in both physical and frequency-space \citep[see][for a detailed review]{D14}.

As Ly$\alpha$ photons diffuse away from the resonance frequency, their mean free path - and hence their escape probability - increases. As a result, the spectrum of Ly$\alpha$ photons emerging from static optically thick media consists of two peaks, which are distributed symmetrically around the resonance frequency (as frequency diffusion can occur to lower and higher energies with equal probability\footnote{For low gas temperatures ($T=10$ K) the energy deposited in the recoiling scattering atom becomes important, and the red peak is enhanced.}). When the scattering medium is contracting [expanding] however, the frequency diffusion preferentially occurs towards higher [lower] frequencies. As a result, the spectrum of Ly$\alpha$ photons emerging from contracting optically thick media is blueshifted. An alternative way to see this is that the converging flow of contracting gas is doing work on the Ly$\alpha$ photons as they scatter outwards, which increases their mean energy \citep{Zheng02,D06a,D06b}.

The Fermi acceleration mechanism described in NM88 invokes similar converging flows of gas. The mechanism is illustrated with a simplified geometry in Figure \ref{setupsFig}, which shows two semi-infinite adjacent slabs of neutral gas. The slab on the left is moving to the right with velocity $v_{\rm s}$ into a stationary slab. If $v_{\rm s} $ exceeds the sound speed in the left slab, then the two slabs are separated by a shock front. 
When a \lya source is in the vicinity of this shock front we expect some \lya photons to diffuse through space and cross the front. When a photon traverses the shock front, the Doppler boost will impart a blueshift in the local gas frame.
Partially coherent scattering off atoms with thermal velocity $v_{th}$ mostly preserves the blueshift of this photon: for a photon with frequency $x_{\rm s} \sim v_{\rm s}/v_{\rm th}$ each scattering event pushes back the photon to the line centre by an average amount $-1/|x_{\rm s}|$ \citep{Osterbrock62}. When the photon scatters back across the shock front before this `restoring force' has returned the photon to line center, the blueshift of the photon increases with each shell crossing. This is what NM88 referred to as `Fermi acceleration'.

In order for Fermi acceleration of a \lya photon to actually occur we require it to scatter after crossing the shock front. The optical depth to a Ly$\alpha$ photon that is blueshifted to $v_b$ through a slab of gas with column density $N_{\rm HI}$ equals $\tau(v_{\rm b})\sim 0.6(N_{\rm HI}/10^{19}\hs{\rm cm}^{-2})(v_b/200\hs{\rm km}\hs {\rm s}^{-1})^{-2}$. The condition $\tau(v_b)\geq 1$ thus translates to $N\geq 1.6\times10^{19} (v_b/200{\rm km}\,{\rm s}^{-1})^2$ cm$^{-2}$. Inversely, a photon can be Fermi accelerated to a maximum blueshift of $v_b \sim 155(N_{\rm HI}/10^{19}\hs{\rm cm}^{-2})$ km s$^{-1}$, {\it in the frame of the scattering medium.} After scattering, the Ly$\alpha$ photon can get an additional Doppler shift $\sim v_{\rm s}$ depending on whether the scattering medium is moving or not.

\section{Fermi Acceleration in Monte-Carlo Simulations}

In this Section we describe the setup and results of the numerical simulations which we use to study the Fermi acceleration mechanism briefly summarised in \S~\ref{mechanism}.

\subsection{Simulation Setup}
\label{simulations}
\begin{figure}
\centering
\includegraphics[width=80mm]{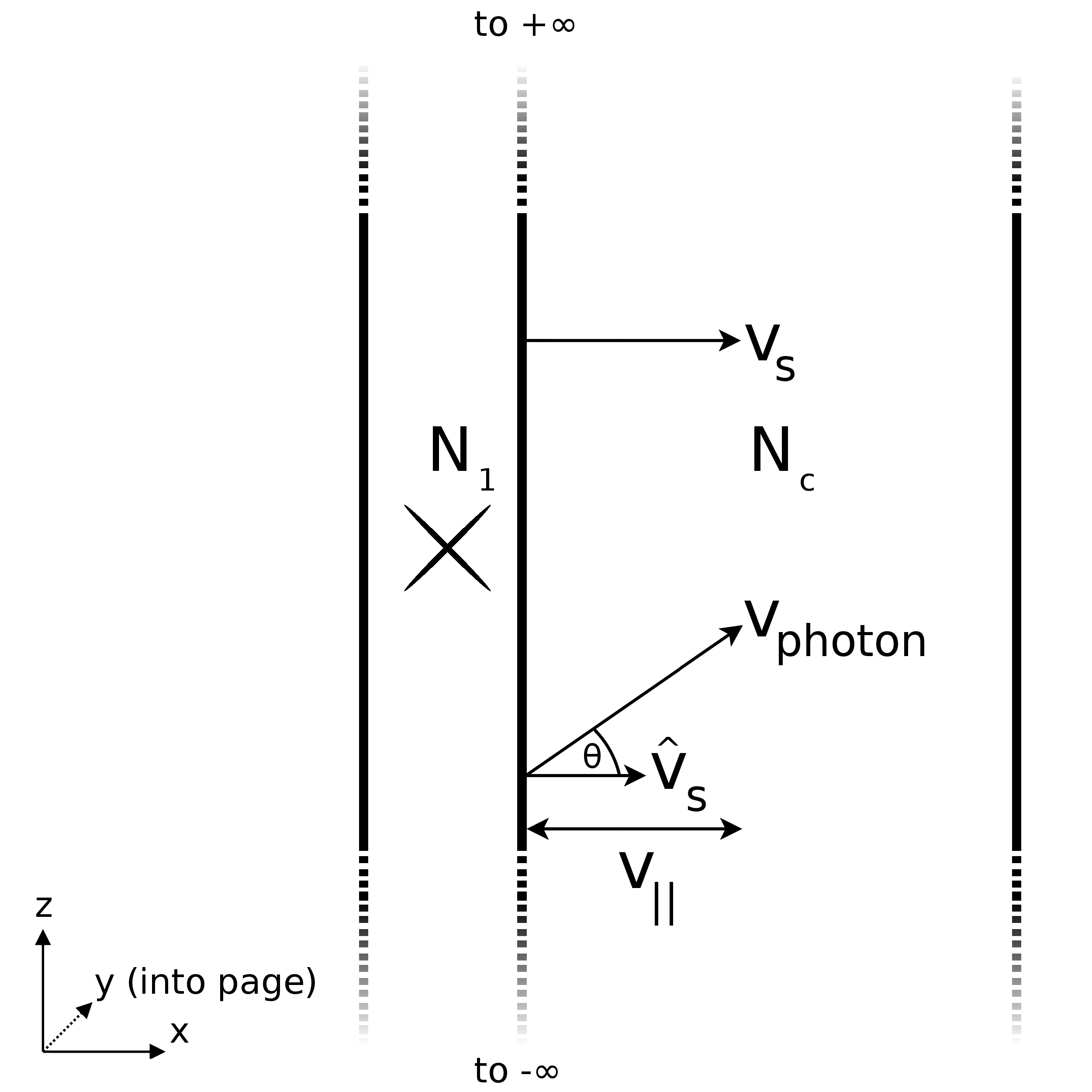}
\caption{Schematic showing the setup used for studying the basic acceleration mechanism (see \S~\ref{mechanism} and \ref{simulations}).  The \lya source is denoted by the cross in the left slab. The left slab has HI column density $N_1$, and the right slab has HI column density $N_{c}$.  The gas in the left slab moves with velocity $v_s$, photons cross the shock front with velocity $v_{photon}$, and the angle between $v_s$ and $v_{photon}$ is labelled $\theta$.  $v_{\parallel}$ is then the projection of $v_{photon}$ onto the unit direction vector of the shock front, $\hat{v}_s$.
}
\label{setupsFig}
\end{figure}

We use modified versions of the \textsc{SLAF} 3D Monte-Carlo \lya radiative transfer code (Chung et al. 2015, in prep), as well as the code described in \citet{D06a}. The vanilla version of \textsc{slaf} is a grid-based code, able to handle arbitrary gas and source distributions within a finite volume of 3D space. For this work we have modified it, removing the dependence on an underlying grid-structure, so that the code can handle the specific setup described in NM88.

We choose to focus on the geometry discussed in NM88 to facilitate a straightforward comparison. Explicitly, there are two distinct regions of gas, represented by slabs of infinite width (y) and height (z).
The slab on the left side represents outflowing gas, and has a bulk velocity of $v_{\rm s} = 400$ km s$^{-1}$ from left to right. The HI column density across the outflowing slab is $N_{1} = 1\times10^{20}$cm$^{-2}$. The slab on the right side represents static gas with an HI column density $N_c= 10^{21}$ cm$^{-2}$ $\gg N_1$. A shock front exists at the interface of the two slabs. \lya photons are emitted at line centre in the rest frame of the outflowing slab on the left, which represents Ly$\alpha$ emission powered by star formation triggered by the passing shockwave (see NM88). We ignore dust throughout the analysis in \S~\ref{simResults}, which simplifies the interpretation of our results and does not affect our main results. We discuss the impact of this assumption separately when we discuss applications of our results in \S~\ref{blueWingsRG} and \S~\ref{blueBumpSpectra}.
Figure \ref{setupsFig} shows the described experimental setup. 

Figure 2 was produced from a run tracking $5\times10^4$ crossing events, while Figures 3 and 4 were produced from runs of $5\times10^4$ photons.
In all cases the results were checked for convergence, and found to already be converged with much lower photon counts.

\subsection{Results}
\label{simResults}

\subsubsection{Shock Crossing Statistics}
\label{crossingStats}
Table \ref{crossingProportions} shows the fraction of photons escaping the simulation volume, $f_{n}$, as a function of the total number of shock crossings, $\emph{n}$. The second and third columns indicate the $f_n$ found in our simulations and the analytic estimate, respectively. The following paragraphs explain how we obtained these analytic estimates.

Photons emitted in the centre of the outflowing slab are equally likely to leave this slab on the left or right hand side. We therefore expect $f_0=1/2$ of all photons not to cross the shock front, which is in agreement with the simulated fraction.

Photons that enter the static slab for the first time are Doppler boosted by an average amount $v_{b}=2v_{\rm s}/3$ (see \S~3.2.2). The probability of these photons being transmitted all the way through the slab is given by $T=4/(3\tau_i)$, where $\tau_i$ denotes the optical depth of the entire static slab to photons that enter at frequency $x_i$ (in the frame of the slab, see \citealt{Neufeld90}; note that this transmission probability only applies when the optical depth of the slab to incoming photons is $\gg 1$).
Substituting numbers yields $T\approx 0.051v^2_{b7}/N^{20}_{\rm HI}$, where $N^{20}_{\rm HI}$ denotes the HI column density of the slab in units of $10^{20}$ cm$^{-2}$, and $v_{b7}$ denotes $v_b$ in units of 100 km s$^{-1}$ (notation adopted from NM88).
We then get $f_1=(1-f_0)\times T\approx 0.018$, which compares reasonably to the fraction $f_1=0.023$ we found in our simulation (exact agreement is not expected as photons cross the slab over a range of frequencies).

Photons that cross the front twice are back in the low column density outflowing slab, and appear Doppler shifted by an average amount $v_{b}\sim 4v_{\rm s}/3$ in the slab frame (see \S~3.2.2).
The slab will appear optically thin to these photons, and we estimate the transmission probability from $T \sim \exp(-2\tau_i)+0.5[1-\exp(-2\tau_i)]$.
Here, $\tau_i=N_1 \sigma_{\alpha}(v_{b})$ denotes the optical depth through the outflowing shell for the incoming photons, where $\sigma_{\alpha}(v_b)$ is the Ly$\alpha$ absorption cross-section at $v_b$.
The factor of $2$ in the exponent accounts for the fact that the photons enter the slab under an angle $\theta$ with probability $P(\cos\theta) \propto \cos \theta$ (see \S~3.2.2). The transmission probability is thus the sum of the probability ({\it i}) that photons are transmitted directly through the slab (the $\exp(-2\tau_i)$ term), and ({\it ii}) that photons that do scatter eventually escape the slab without crossing the shock again.  For the latter, we assume that photons are equally likely to escape on the left and right hand side after scattering, which seems reasonable given that the slab appears optically thin on average.  Substituting numbers we obtain $f_2=(1-f_0-f_1)T \sim 0.28$, which agrees with the simulation result.

With $T=4/(3\tau_i)$ and $v_b \sim 2v_{\rm s}$, we estimate $f_3=(1-f_0-f_1-f_2)\times T \sim 0.068$, 48.5\% larger than what we obtain from the simulation. We attribute this discrepancy to the approximation $T=4/(3\tau_i)$ breaking down at larger shifts from line centre.

The previous analysis allows us to understand quantitatively the simulation results. More photons escape after undergoing an even number of shock crossings than odd. This is simply a reflection of the fact that in our experimental setup we have specified that $N_c \gg N_1$. 

As we pointed out in \S~\ref{introduction}, it is theoretically possible that a \lya photon that has been Fermi accelerated into the line wing can subsequently diffuse back into the core of the line prior to escaping or crossing the shock front again. When this happens, any memory of previous Fermi acceleration is erased. However, our numerical simulations indicate that in practice this almost never occurs, and therefore we can identify odd [even] numbered contiguous shock crossings with photons which exit on the right [left] side of the simulation.

\begin{table}
\caption{Fraction of photons undergoing \emph{n} successive shock crossings before exiting the simulation (middle column). The right column refers to the same fraction calculated with an analytic method.}
\centering
\begin{tabular}{|c|c|c|}
  \hline
  total \# shock crossings ($n$) & fraction ($f_n$)& analytic estimate\\
  \hline
  0 & 0.50820 & 0.5 \\
  1 & 0.02299 & 0.018\\
  2 & 0.26215& 0.28\\
  3 & 0.04578 & 0.068 \\
  4 & 0.12525 &...\\
  5 & 0.01283 &...\\
  6 & 0.01944 &...\\
  7 & 0.00149 &...\\
  8 & 0.00168 &...\\
  9 & 0.00008 &...\\
  10 & 0.00011 &...\\
  \hline
\label{crossingProportions}
\end{tabular}
\end{table}

\subsubsection{Velocity Shift vs Shock Crossing Number}
\label{sec:velshift}

We denote the angle at which a photon crosses the shock front with $\theta$ (see Fig~\ref{setupsFig}). For an isotropic distribution of shock crossing directions we have $P(\mu)=1$, where $\mu \equiv \cos \theta$. The average Doppler shift experienced by a photon as it crosses the shock front is $v_{b}=v_s \int_0^1d\mu\hspace{1mm} \mu P(\mu)=v_s/2$. After $l$ crossings the photon experiences an average blueshift of $lv_s/2$ (NM88).

\begin{figure}
\centering
\includegraphics[width=80mm]{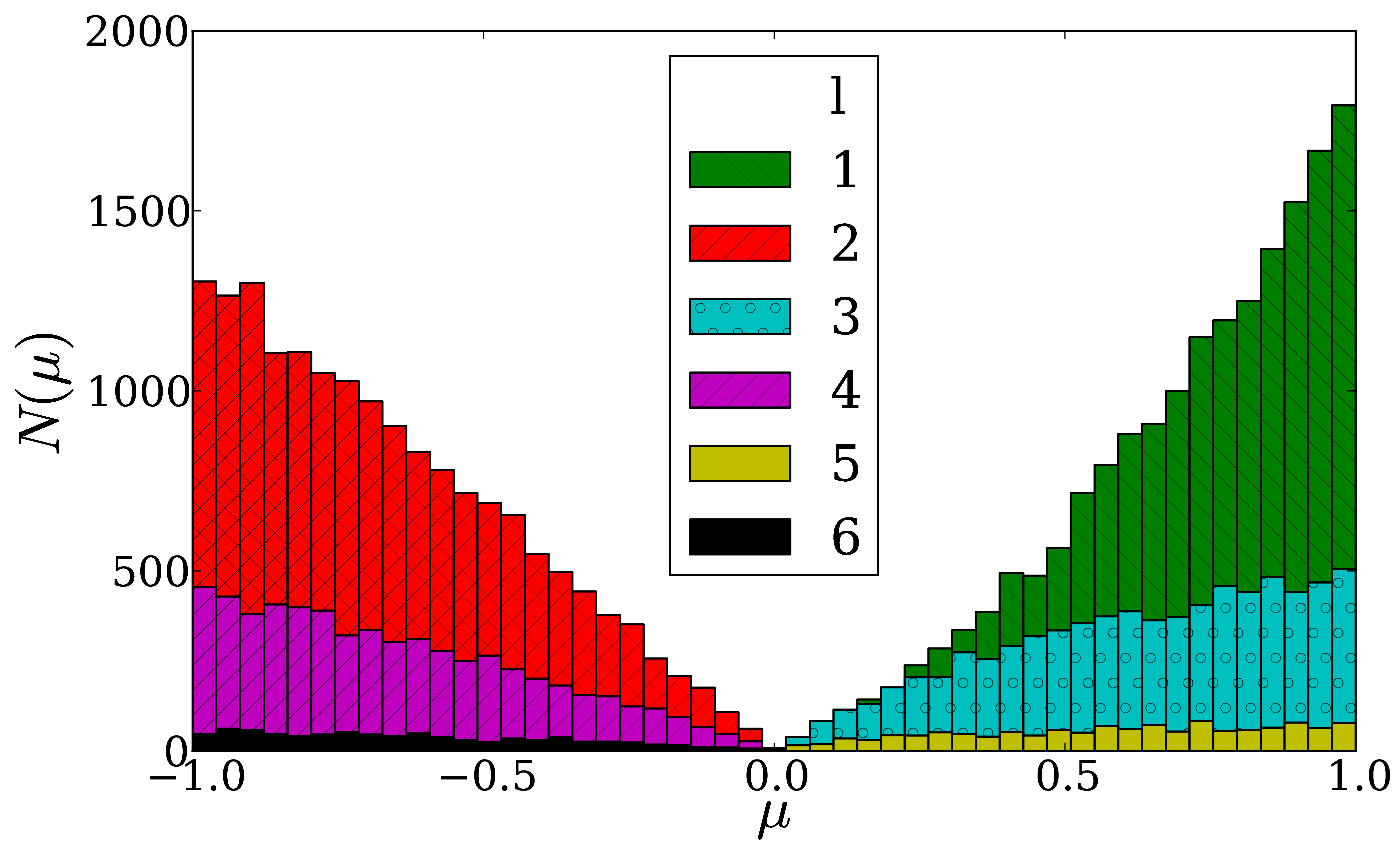}
\caption{Distribution of the projection of the shock-crossing angle for various values of $l$.  For each value of $l$ the plotted distribution includes all photons that make an $l^{th}$ crossing regardless of what their final crossing count, $n$, is.}
\label{crossingDistr}
\end{figure}

With our code we are able to track individual photons as they cross the shock front, which allows us to directly measure $P(\mu)$.
Figure \ref{crossingDistr} shows the number count, $N(\mu)$, for photons that cross the $l^{th}$ time for various $l$. $P(\mu)$ only differs from $N(\mu)$ by a normalisation factor.  Here, $l$ refers to the current (rather than final) contiguous crossing count of a particular photon as it crosses the shock front. A photon which finally exits the simulation after $n$ crossings is represented $n$ times (once for each crossing). Figure \ref{crossingDistr} shows clearly that the distribution of photon crossing projections is not isotropic\footnote{In Figure \ref{crossingProportions} $\mu>0$ [$\mu <0$] for photons that cross the shock front from left-to-right [right-to-left]. In Figure \ref{crossingDistr} only photons on an odd [even] shock crossing contribute to $N(\mu)$ for positive [negative] values of $\mu$.}.
Instead, we find that $P(\mu) \propto \mu$. This distribution has been found in previous analyses \citep[e.g.][]{Ahn01,Nicolas14}, and its origin is discussed in Appendix~\ref{app:pmu}. For this distribution we expect $v_s \int_0^1d\mu \hspace{1mm} \mu P(\mu)=2v_s/3$ (see also Fig.~\ref{projVsScattering}), and hence that a photon experiences a Doppler boost $2l v_s /3$ after $l$ shock crossings\footnote{Note that if a photon crosses the shock front at an angle $\theta$, then the total hydrogen column density to the edge of the slab is $N_{\rm HI}/\mu$, where $N_{\rm HI}=N_1$ [$N_{\rm HI}=N_c$] for the slab on the left [right] in Figure~\ref{setupsFig}. The angle-averaged column to the edge of the slab is given by $\langle N_{\rm HI} \rangle = \int_0^1 d\mu \hspace{1mm}P(\mu)N_{\rm HI}/\mu=2N_{\rm HI}$, as used in \S~3.2.1.}.

\begin{figure}
\includegraphics[width=85mm]{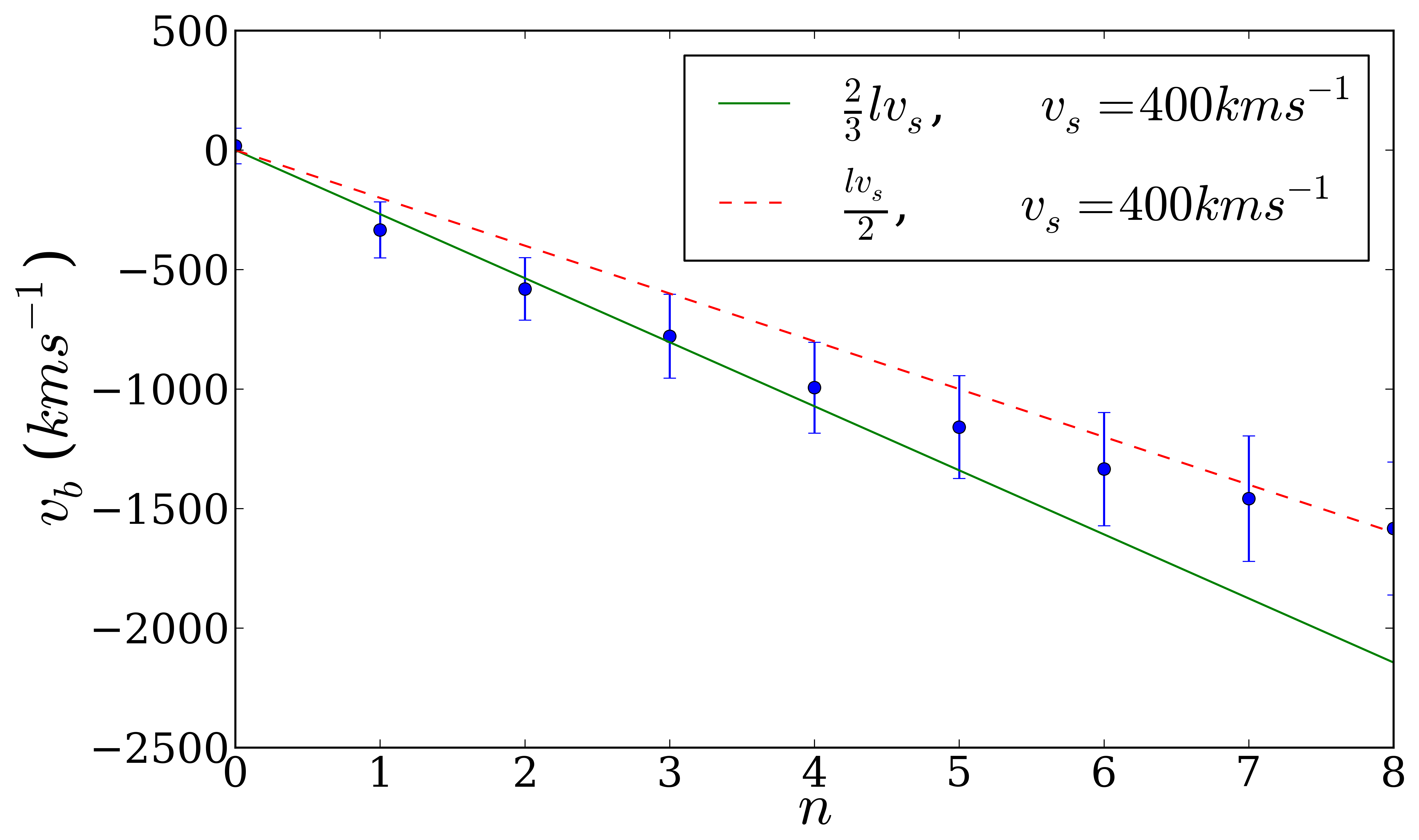}
\caption{Mean velocity offset for photons exiting with $n$ shock crossings as found in our simulations (points), as measured in the frame of the gas.  The error bars show the standard deviation within the distribution.  The dashed red line shows the $\frac{1}{2}lv_s$-relation which applies to isotropic shock crossings, whereas the green line refers to the $\frac{2}{3}lv_s$-relation which applies to anisotropic shock-crossings of the form $P(\mu) \propto \mu$ (see text).}
\label{velShiftVsScatterings}
\end{figure}

Figure \ref{velShiftVsScatterings} shows the velocity shift $v_b$ of exiting photons as a function of the total number of shell crossings $n$, in the frame of the gas into which the photon is crossing (i.e. for odd [even] numbered shock crossings the velocity shift is given in the static [outflowing] frame). Results from our Monte-Carlo simulations are represented by the data points. The red dashed line shows the analytic result under the assumption that photons cross the shock front isotropically \citep[as in][]{McKee}, while the green solid line shows the analytic result for photons crossing the shock front according to $P(\mu) \propto \mu$. For $n \leq 3$ the simulation results follow the $\frac{2lv_s}{3}$-relation (where here $n$ is a good proxy for $l$), after which it approaches the isotropic shock-crossing case. The reason for this transition is that as the photons get blueshifted further into the wing of the line profile their mean free path increases. As we discuss in detail in Appendix~\ref{app:pmu}, we expect a transition to isotropic shock crossing when the mean free path becomes comparable to the thickness of the slab.\\

\begin{figure}
\includegraphics[width=80mm]{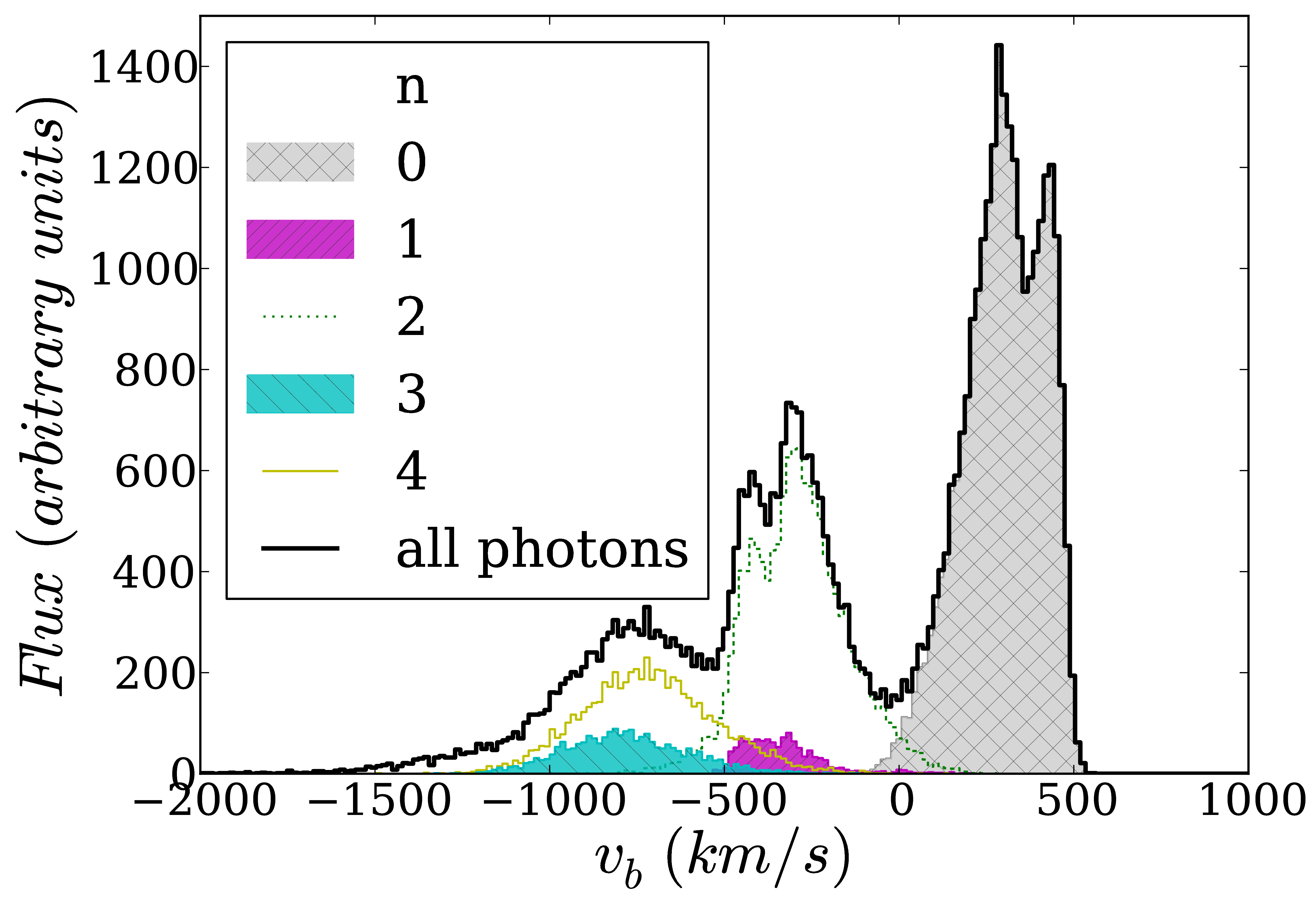}
\caption{Spectrum of photons exiting the slabs with a single source, lab frame. $n$ is the number of shock crossings that the photon undergoes prior to exiting the simulation (see \S~\ref{crossingStats}).}
\label{spectrumSingleSource}
\end{figure}

Finally, Figure \ref{spectrumSingleSource} shows the spectrum outcoming from the slab, as measured in the lab frame. The result can be easily understood in terms of our previous analysis. More specifically, the total spectrum (black histogram), can be explained as follows.
\begin{itemize}
\item The peak at $v_{b} \sim 300$ km s$^{-1}$ is composed primarily of the $f_0=50\%$ of all photons that did not cross the shock front. These photons diffused outward of the left slab, and would have emerged with a characteristic double peaked emission line profile centered around the resonance frequency \citep[e.g.][]{Adams1972,Harrington1973,Neufeld90}. However, when we Doppler boost these photons back into the lab frame this double peak is diluted by the fact that the photons escaping the slab do so at various angles.  The Doppler boost is dependent on the exit angle, with an average Doppler boost of $\langle \mu \rangle v_{\rm s} \sim 270$ km s$^{-1}$. The grey crosshatched histogram shows the spectrum of these photons.
\item The small fraction of photons that escape after a single shock crossing, $f_1\sim 2\%$, is indicated by the purple hatch-filled histogram and escapes with a mean blueshift of $\sim 2v_s/3 \sim 300$ km s$^{-1}$ (also see Fig.~\ref{velShiftVsScatterings}, these photons escape from the static shell, and no additional Doppler boost into the lab frame is required). 
\item The photons that escape after two shock crossings (green dotted line histogram) account for $f_2 \sim 26\%$ of the total. Figure~\ref{velShiftVsScatterings} shows that these photons have accumulated a blueshift of $\sim 4v_s/3 \sim 530$ km s$^{-1}$ in the frame of the outflowing gas. A Doppler boost back into the lab-frame transforms these photons back to a blueshift of  $\sim 2v_s/3 \sim 300$ km s$^{-1}$. 
\end{itemize}

A similar reasoning applies to photons that escape after a larger number of crossings, and shows that it is easy to understand the shape of the total spectrum: the redshifted peak (grey crosshatched histogram) at $v_b \sim 2v_s/3$ is composed of photons that did not cross the shock front. The first peak blueward ($v_b<0$) of the \lya resonance is at $v_b \sim -2v_s/3$ and consists of photons that crossed the shock front $1-2$ times. Similarly, the second blueward peak at $v_b \sim -4v_s/3$ consists of photons that crossed the shock 3-4 times, etc. The prevalence of these peaks depends on a number of factors, including $v_s$, $N_1$, $N_c$, and also the distribution of Ly$\alpha$ sources relative to the HI gas.

Finally we note that our results here are in excellent agreement with the analytic estimate of the maximum blueshift detailed in \S~\ref{mechanism}.
Setting $N=10^{21}$\,cm$^{-2}$ in the analytic formula gives $v_b \approx 1600$ km s$^{-1}$, which lies in the blue tail of Figure \ref{spectrumSingleSource}.

\section{Applications}
\label{applications} 

We now explore, by means of our Monte-Carlo codes and assuming some simple models, whether the Fermi acceleration mechanism can provide a viable explanation for some observed systems.

\subsection{Extended Blue Wings in Radio Galaxies}\label{blueWingsRG}

The analysis by NM88 was motivated by observations of 3C326.1, a radio source with a Ly$\alpha$ spectral line profile with emission extending far into the blue wing of the line profile (\citealt{1988ASSL..141..259D}, \citealt{1987hrpg.work...85S}).
NM88 modelled 3C326.1 with a shock front that was propagating into a collection of dense clumps.
The shock passage triggers star formation and Ly$\alpha$ emission inside the post-shock gas.
The subsequent radiative processes in each clump proceed just as in the slab models we discussed previously.
The details of the emerging spectrum depend on the distribution and covering factor of the clumps, but for simplicity and to facilitate a direct comparison to NM88 we follow their prescription and assume that the emerging spectrum is a superposition of the spectra of individual clumps. NM88 further show that dust does not affect the emerging spectra, provided that the gas-to-dust ratio $\xi_{\rm dust}\lsim 0.016$, where $\xi_{\rm dust}$ denotes the dust-to-gas ratio relative to the local interstellar value. They argue that this limit is acceptable because the dust grains would be destroyed by the passage of the shock front. We finally assume all clumps - and therefore the radiative transfer processes inside them - to be identical, then we can simply adopt the spectrum from Figure~\ref{spectrumSingleSource}. In reality, from our findings in \S~3.2.2, we expect that a distribution of clump properties would give rise to a superposition of spectra with peaks in different locations, and therefore that the emission in the blue wing of the line profile would have less prominent features\footnote{Even for fixed clump properties, the emerging Ly$\alpha$ spectrum depends on the placement of Ly$\alpha$ sources.  In particular, changing only the source position leaves the peak positions unchanged, but modulates their relative amplitudes.  We have explicitly verified, however, that our spectrum barely changed if we assumed a uniform distribution of Ly$\alpha$ sources throughout the outflowing slab.}.

Putting these reservations aside, we observe that qualitatively our simulated spectrum shows similarities to that predicted by NM88.
We should note though that in Figure~\ref{spectrumSingleSource} we omit the direct sources at $v_b=0$, which NM88 put in by hand as a delta function at $v_b=0$.

We note that the two largest peaks immediately redward and blueward of line centre cannot be attributed to photons exiting without crossing the shock front as in NM88 (i.e. the standard double-peak profile produced by a static, optically thick medium).
Instead, our analysis shows clearly that these peaks are primarily caused by photons which have undergone 0 and \{1, 2\} shock crossings respectively. The smaller, secondary blueward peak is primarily composed of photons which have undergone 3 or 4 shock crossings.
The primary differences between the spectra predicted by us and by NM88 are in the asymmetry of the two main peaks, and the ratio of the peaks. As we have previously mentioned, the source distribution and physical parameters of the clumps affect the outcoming spectrum, and it is likely that tuning of either or both of these will alleviate this discrepancy.
The two blue peaks in our model appear to be shifted redward compared to the model predictions from NM88.
However, a comparison of the observed spectrum in NM88 and our model shows that the positions of our blue peaks are also a good fit to the data.
In fact, if the amplitude of the peaks is ignored (this is dependent on the exact source distribution), the position of the peaks in our model is arguably a better fit to the data than the NM88 model.

We now consider the observed Ly$\alpha$ spectra of $z\sim 3.4$ radio galaxy B2 0902+34 \citep{Adams}.
There is broad agreement in the shape of the observed and predicted spectra, with the broad wing extending well beyond $\sim 1000$ km s$^{-1}$, and some spectra displaying prominent peaks blueshifted by $\sim 1000$ km s$^{-1}$.
The spectra from B2 0902+34 vary widely depending on where exactly on the galaxy the fiber is placed.
It is however true to say that for the majority of fibers the most prominent peak has been observed to lie either at the systemic velocity, which was determined from the observed 21-cm absorption signature, or slightly blueward of it.
This would appear to be in tension with our model here which, following NM88, predicts the most prominent peak to appear at $v_b=0$ (not shown in Figure \ref{spectrumSingleSource}) as a result of direct `blister' sources on the edge of the clump.
Recalling that the amplitude ratios of the peaks is determined by the source distribution, we speculate that this tension could be relieved if the most prominent peak in the B2 0902+34 observations is instead identified as the first blueshifted peak in our model, and the absence of a pronounced peak at systemic velocity attributed to a lower prevalence of `blister' sources in this system.

\subsection{Blue Bump Spectra}\label{blueBumpSpectra}

As we mentioned in \S~\ref{introduction}, observed Ly$\alpha$ spectral line profiles can often be reproduced surprisingly well with shell-models. In these models, a central Ly$\alpha$ source is surrounded by a geometrically thin shell of gas. The shell-models contain two parameters describing the Ly$\alpha$ source: ({\it i}) the assumed FWHM of the Ly$\alpha$ line prior to scattering, and ({\it ii}) the strength of the Ly$\alpha$ emission line, which is quantified by the equivalent width (EW). The shell itself is described by four additional parameters: ({\it i}) the HI column density of the shell $N^{\rm shell}_{\rm HI}$, ({\it ii}) its outflow velocity $v_{\rm shell}$, ({\it iii}) its dust content $\tau_{\rm d}$, and ({\it iv}) its velocity dispersion $b$.

Some recent analyses have pointed out that shell-models have difficulties explaining blue bumps in observed spectra \citep[e.g.][]{Kulas12,Chonis13,Hashimoto15}. Nice examples can be seen e.g. in Figure~7 of \citet{Chonis13}. More recently \citet{Hashimoto15} pointed out that the shell-models, when applied to blue-bump objects, consistently require the intrinsic Ly$\alpha$ line to have a FWHM which is too large (an excessively high FWHM was only required for blue-bump objects).

Because Fermi acceleration naturally gives rise to blueshifted emission, we investigate a minor modification of the shell-model in which we embed the outflowing shell within a static gas cloud. This modification can be interpreted as the situation in which the outflow is still propagating into the static interstellar medium (possibly prior to breaking out of it). This configuration now includes a shock front as in our previous analysis. We specifically study a shell-model with parameters based loosely on those inferred by \citet{Hashimoto15}: $N^{\rm shell}_{\rm HI}=10^{19}$ cm$^{-2}$, $v_{\rm shell}=200$ km s$^{-1}$, and FWHM=200 km s$^{-1}$. We further assume $\tau_{\rm d}=0$, EW$=\infty$ (i.e. pure \lya emission), and $b=12.9$ km s$^{-1}$ (corresponding to gas at $10^4$K), but note that these assumptions do not affect our results at all\footnote{For completeness we have presented our numerical results which include dust in Appendix~\ref{app:dust}. These results show clearly that dust barely affects the blue bumps. This result is not surprising: our modification only adds a small amount of additional hydrogen, and this additional gas triggers Fermi-acceleration into the (blue) wings of the Ly$\alpha$ line profile, where  \lya photons escape more easily.}. Finally, the key new model ingredient is the addition of a static shell of gas adjacent to the outflowing shell. This shell is characterised primarily by its column density $N_{\rm HI}^{\rm stat}$. We assume that it has no dust and that it has the same temperature as the outflowing shell.

\begin{figure}
\includegraphics[width=80mm, angle=0]{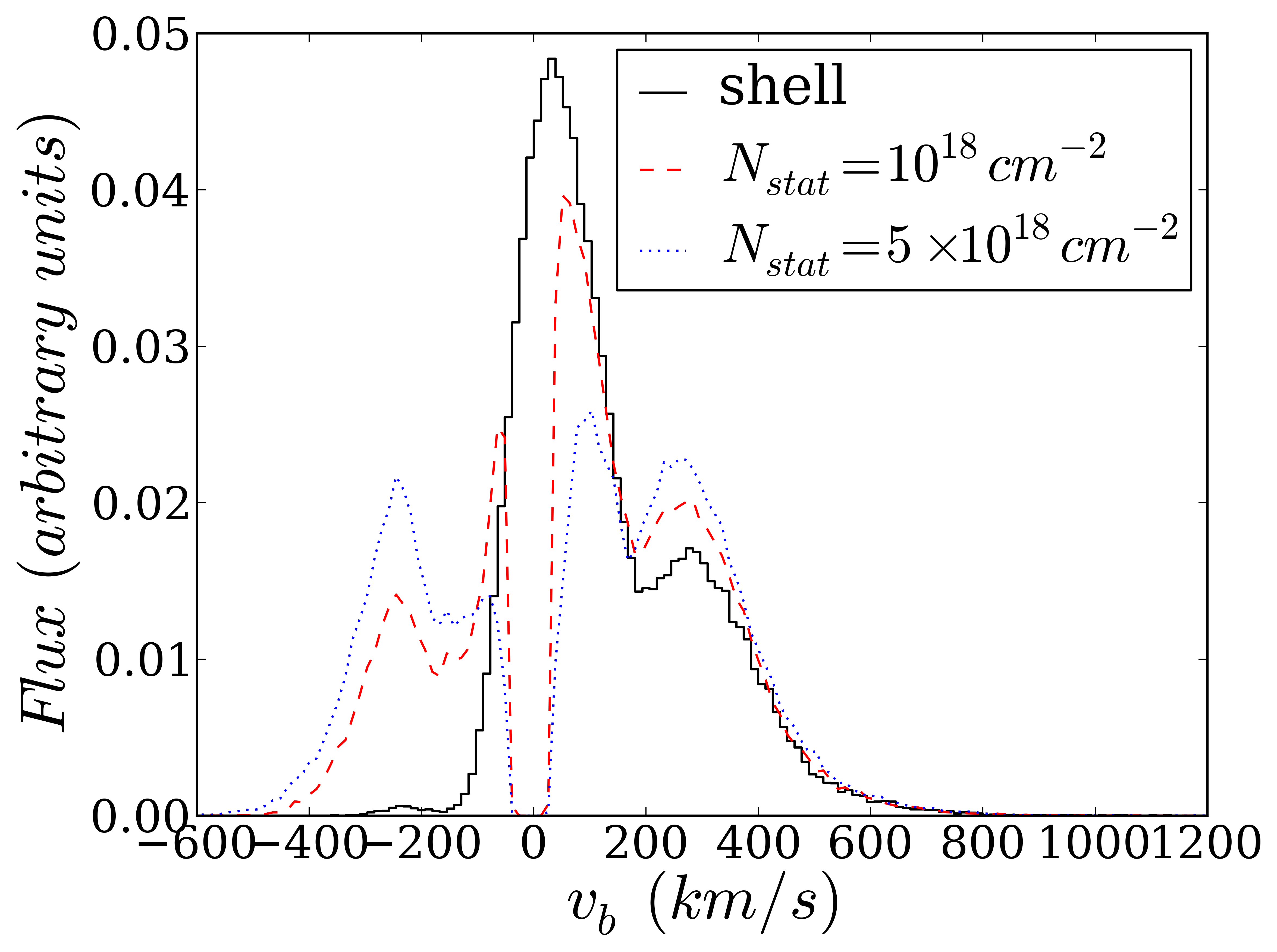}
\caption{Predicted Ly$\alpha$ spectra emerging from a Ly$\alpha$ source surrounded by a shell of HI gas outflowing at 200~km~s$^{-1}$ (the shell-model, see text) embedded within a static neutral shell with a column density $N_{\rm HI}^{\rm stat}$. The solid black histogram shows the spectrum for $N_{\rm HI}^{\rm stat}=0$ (i.e. the standard shell-model), while the red dashed and blue dotted line shows the spectrum for $N_{\rm HI}^{\rm stat}=10^{18}$ cm$^{-2}$ and $N_{\rm HI}^{\rm stat}=5\times10^{18}$ cm$^{-2}$, respectively. The plot demonstrates that adding even a small amount of hydrogen (1/10$^{\rm th}$ of that in the shell) triggers the on-set of Fermi acceleration, which gives rise to a blue bump.}
\label{fig:shellresults}
\end{figure}

\begin{figure}
\includegraphics[width=85mm]{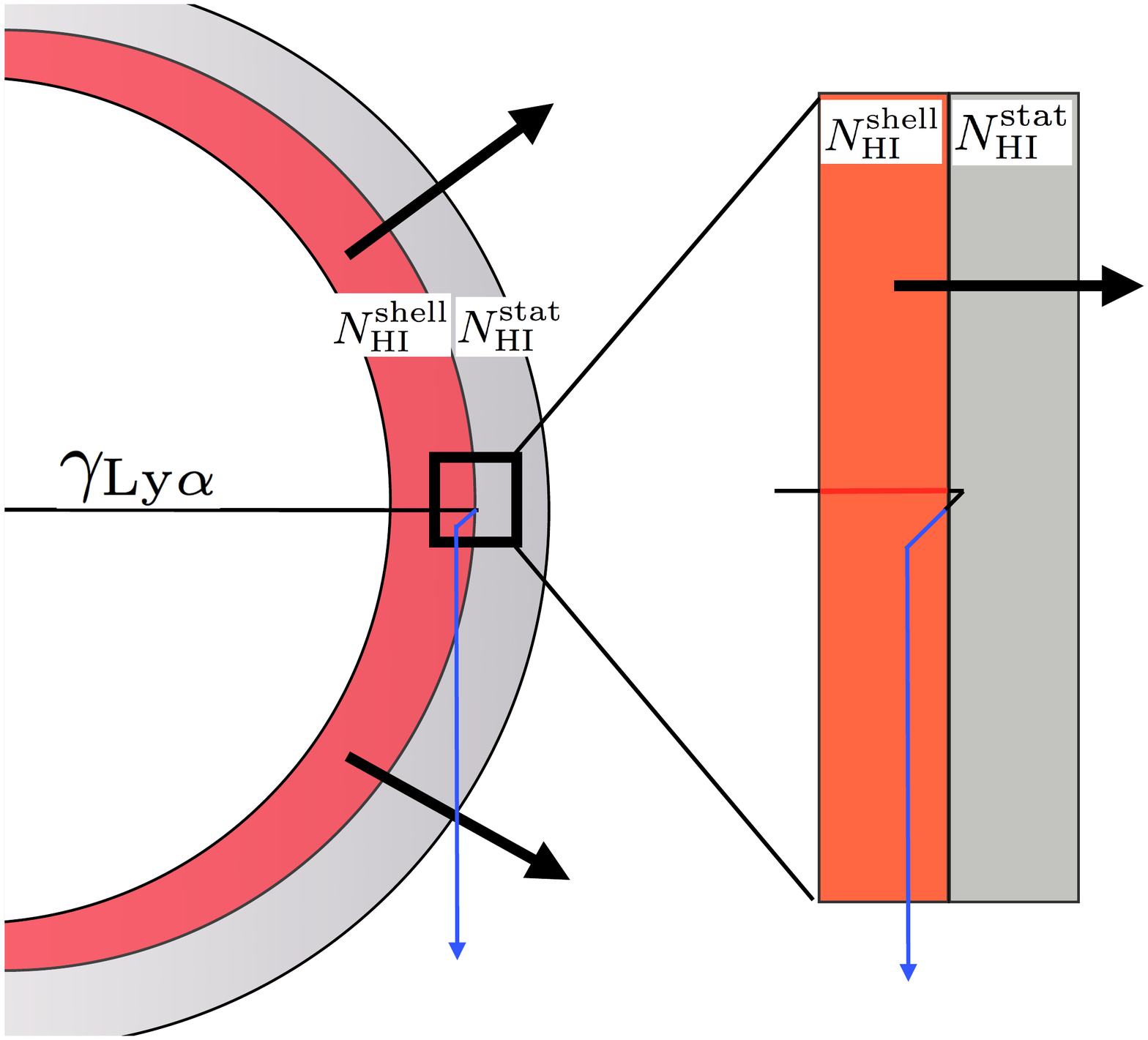}
\caption{Schematic illustration of the origin of photons of the blue bump observed in Ly$\alpha$ spectra in the shell-model. The outflowing shell is surrounded by a static ISM. }
\label{fig:shell}
\end{figure}

The results of this analysis for $N_{\rm HI}^{\rm stat}=0$ (i.e. the original shell-model; solid black histogram), $N_{\rm HI}^{\rm stat}=10^{18}$ cm$^{-2}$ (red dashed line), and $N_{\rm HI}^{\rm stat}=5\times 10^{18}$ cm$^{-2}$ (blue dotted line) are shown in Figure~\ref{fig:shellresults}, 
where it is clear that a small additional column of hydrogen ($N_{\rm HI}^{\rm stat} \ll N^{\rm shell}_{\rm HI}$) dramatically affects the spectrum blueward of the systemic velocity. This large change of the spectrum can be easily understood, as the outflowing shell with $N^{\rm shell}_{\rm HI}=10^{19}$ cm$^{-2}$ directly transmits a significant fraction of Ly$\alpha$ photons. However, the surrounding static shell is optically thick to photons emitted near line centre, because these photons still appear close to the centre of the line in the frame of this gas. The static shell therefore effectively reflects back photons into the outflowing shell\footnote{In Figure~\ref{fig:B2} we show that this mechanism produces blue bumps for column densities as low as $N_{\rm HI}^{\rm stat}\sim 10^{15}$ cm$^{-2}$. This is because a static shell of gas with $N_{\rm HI}^{\rm stat}\sim 10^{15}$ cm$^{-2}$ remains optically thick to Ly$\alpha$ photons emitted close to line centre, and can therefore reflect back photons into the expanding shell.}. The reflected photons appear blueshifted by $\sim v_{\rm shell}$ in the frame of the outflowing shell, where their newly acquired large blueshift makes them escape efficiently. For example, photons that are scattered by $90^{\circ}$ after being reflected back into the outflowing shell escape with a blueshift of $\sim v_b$, which is indeed where the new peaks in the spectrum emerge. This process is depicted schematically in Figure~\ref{fig:shell}. The analytic estimate for the maximum blueshift in \S~\ref{mechanism} provides a decent estimate of the maximum blueshift $v_b\sim 160 {\rm km\,s}^{-1}+ v_s \sim 350$ km s$^{-1}$ (where the $160$ km s$^{-1}$ was the maximum shift for a column density of $10^{19}$ cm$^{-2}$ in the frame of the scattering medium).

\section{Discussion and Conclusions}\label{conclusions}
We have presented a detailed analysis of Fermi acceleration of Ly$\alpha$ photons across a shock front, a process that was first studied analytically by NM88.
Our Monte-Carlo radiative transfer simulations of the slab model (Figure \ref{setupsFig}) confirm the basic results in NM88. 

In particular, we find that each time a photon crosses a shock front that is propagating at $v_s$, it experiences a Doppler boost $v_b\sim (0.5-0.7) v_s$ in the gas frame. The precise numerical coefficient depends on the opacity of the gas to the Ly$\alpha$ photons. We also find the blueshift of the Ly$\alpha$ photons to increase in proportion with the number of shock crossings $l$, i.e. $v_b \propto (0.5-0.7) lv_s$.

We discussed how our results can help to explain extended blue wings observed in spectra of radio galaxy 3C326.1 (which was proposed previously by NM88), but we note that our line as a whole appears redshifted by $\sim 300$ km s$^{-1}$. We reached the same conclusions for the radio galaxy B2 0902+34 \citep{Adams}. We nevertheless consider Fermi acceleration a plausible alternative to the model proposed by \citet{Adams}, which involves the collapse of $> 10^{12} M_{\odot}$ of neutral gas.

Last, we show that Fermi acceleration naturally gives rise to blue bumps in Ly$\alpha$ spectra, which are difficult to reproduce with conventional shell-models (at least those with reasonable values for the FWHM of the intrinsic Ly$\alpha$ line, see \citealt{Hashimoto15}). We presented a natural extension to shell-models in which the shell is expanding into static gas that contains a low column density of HI\footnote{If we interpret the shell as dense gas that is swept up by feedback processes, then the static gas can be interpreted as gas that is being swept up.}. These models can give rise to blue bumps in Ly$\alpha$ spectra without abandoning the simplicity of the conventional shell-model\footnote{It is possible that this simply signals the break-down of the shell-model, and that radiative transfer through more complex gas geometries needs to be studied. There are numerous works studying Ly$\alpha$ transfer through more complex, simulated gas distributions \citep[e.g.][]{Tasitsiomi2006,Laursen2009,Barnes11,Verhamme12,Behrens14,Lake15,Yajima2015}, though results from these calculations have not been compared to observed spectra in detail. Note however, that the sensitivity of the Ly$\alpha$ transfer scattering process to the distribution and kinematics of neutral gas implies that these predictions depend on sub-grid feedback prescriptions as well as the spatial resolution of the simulation (which ideally needs to be sub-pc, see Dijkstra \& Kramer 2012 for a discussion). Simulations therefore have their own uncertainties, and thus provide a complementary route to addressing Ly$\alpha$ scattering on interstellar scales.}. In our model, the bump consists of photons that initially streamed through the outflowing shell, but which were reflected back into this shell by the surrounding static interstellar medium. This suggests that blue bumps in Ly$\alpha$ spectra are associated with outflows that are still confined to the ISM of the galaxy, which may represent an earlier stage in the evolution of the galaxy.

\section*{Acknowledgments}
We thank Koki Kakiichi and Michele Sasdelli for enlightening discussions, and the referee for critical, helpful comments.

\bibliographystyle{mn2e}

\appendix
\section{Origin of $P(\mu) \propto \mu$}
\label{app:pmu}

The goal of this appendix is to clarify why the distribution of angles at which Ly$\alpha$ photons cross the shock front scales as $P(\mu) \propto \mu$ (where $\mu \equiv \cos \theta$), rather than isotropically.

Photons that cross the shock front will on average have traversed an optical depth $\tau=1$.
Depending on $\mu$, $\tau=1$ corresponds to a different (frequency dependent) physical depth, $d(\nu)$, away from the shock front.
It is easy to see that $d \propto \mu$, so that at fixed frequency the volume of gas that the photon is likely to have last scattered in is $V_{\rm scat} \propto \mu$.
If we further assume that the density of scattering events is homogeneously distributed in the gas, then this leads directly to the relation $I \propto \mu$, where $I$ is the intensity of photons.
Therefore $P(\mu)\propto \mu$, which we showed in Figure \ref{crossingDistr}, and which was found previously by \citet{Ahn01} and \citet{Nicolas14}.

\begin{figure}
\includegraphics[width=80mm]{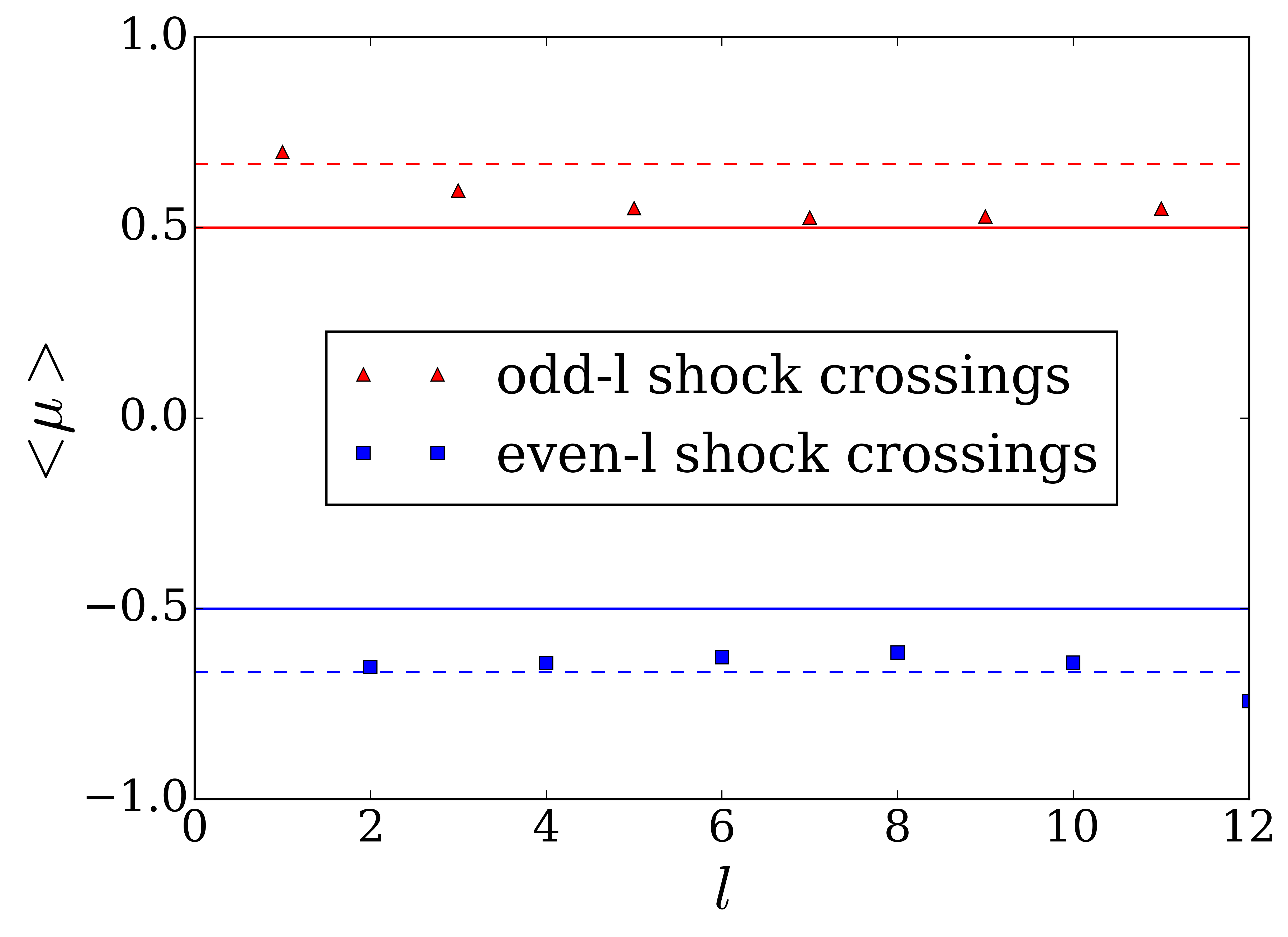}
\caption{Mean projection as a function of shock crossing number, $l$.  The solid lines represent the expected values for isotropic scattering, while the dashed lines show $<\mu>=\pm2/3$.  Red triangular points are for odd-$l$ shock crossings, blue square points are for even-$l$ shock crossings.
}
\label{projVsScattering}
\end{figure}

Figure~\ref{projVsScattering} shows the average angle at which photons cross the shock front, $\langle \mu \rangle =\int_0^1 d\mu\hspace{1mm}\mu P(\mu)$, as a function of crossing number $l$.
As we mentioned in the paper, odd [even] number of shock crossings correspond to crossings from left-to-right [right-to-left].
This Figure clearly shows that $\langle \mu \rangle \sim 2/3$ for even shock-crossings, while it approaches $\langle \mu \rangle \sim 0.5$ for odd-shock crossings when $l > 3$. 

\begin{figure}
\includegraphics[width=80mm]{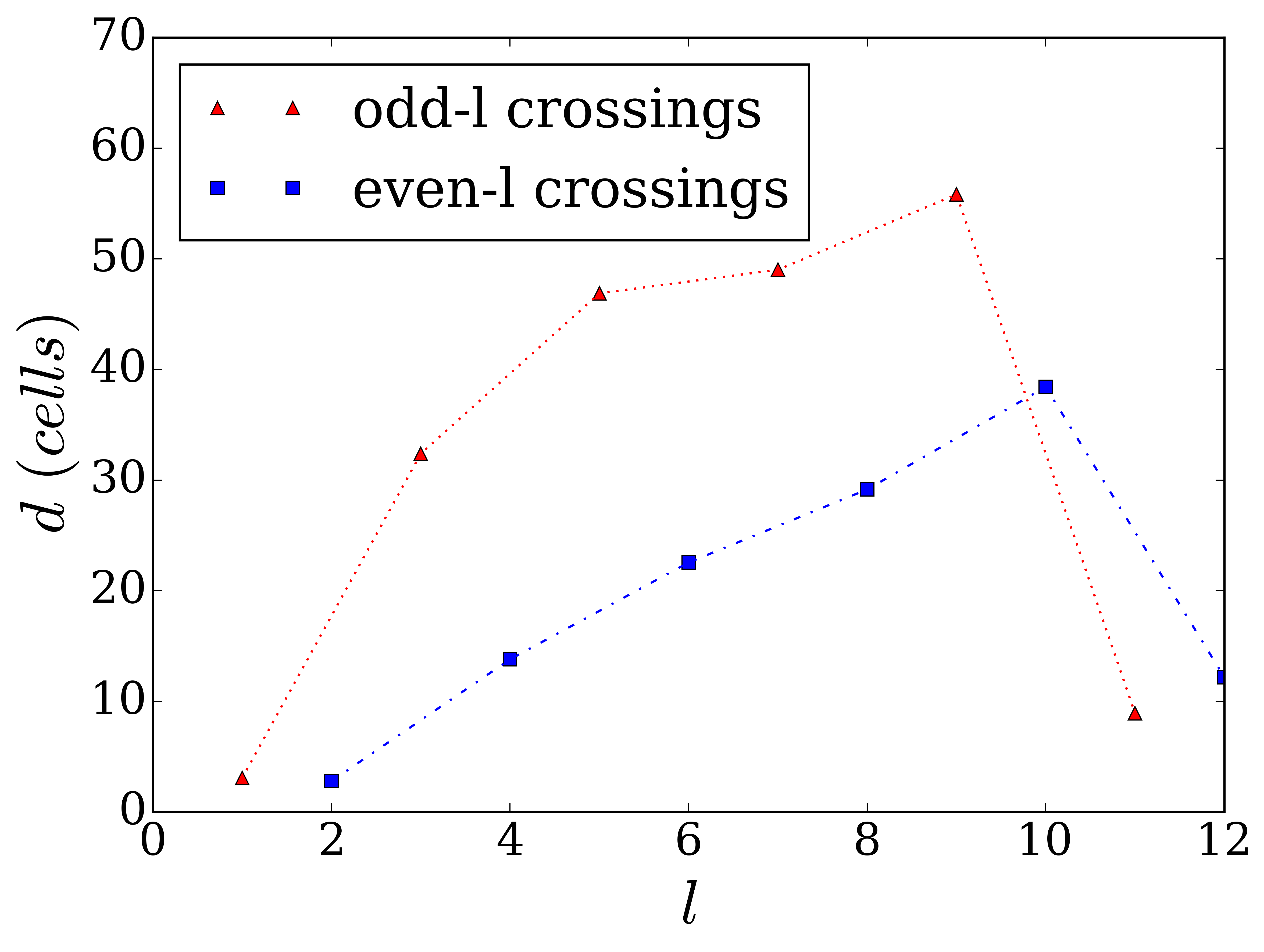}
\caption{Mean distance between the shock and the location of the last scattering before crossing the shock vs shock crossing number, $l$.  d is shown in simulation units, i.e. cells (each slab is 128 cells wide). Red triangular points and the red dotted line show the odd-l crossings, while blue square points and the blue dot-dash line show the even-l crossings.}
\label{depthFromShock}
\end{figure}

An explanation for this is offered in Figure \ref{depthFromShock}, which shows the mean distance from the shock front (scattering depth) of the last scattering before crossing the shock as a function of the shock crossing number $l$.
This distance follows an inverse exponential distribution.
As the photon becomes more blueshifted with increasing $l$, the mean free path of the photon also increases, so that for each successive $l$ the photon is coming from deeper within the slab, explaining the increasing trend in Figure \ref{depthFromShock}.
However, as the mean free path increases, the photons which propagate the farthest actually propagate all the way through the gas slab and exit the simulation.
Therefore, the distribution of scattering depth is truncated for the next scattering count, resulting in a flattening in the growth of scattering depth as a function of $l$.
This is precisely the behaviour we observe for odd-$l$ in Figure \ref{depthFromShock}.
The scattering depth distribution is truncated, but the photon blueshift is unaffected.
Therefore, the gas between the next scattering event and the shock front becomes effectively optically thin to the photon, and we expect photons to cross the shock front isotropically, which is why $\langle \mu \rangle$ approaches $0.5$ in Figure \ref{projVsScattering}.

For even-$l$ we do not observe this transition to isotropic crossing.
This is because $N_1 \ll N_c$, so that at a given $l$ and blueshift the mean free path of a photon is much lower in the right (static) slab and the scattering depth distribution for even-$l$ (i.e. crossings where the previous scattering was in the right, static slab) has not yet become truncated.
As discussed above it is the truncation of the scattering depth distribution that causes the transition to isotropic shock crossing, and so in its absence for even-$l$ the Ly$\alpha$ photons still cross the shock front following $P(\mu) \propto \mu$.

Finally, we note that the highest shock crossing counts that occur in our simulation are rare events, and the resultant poor statistics are responsible for the turnaround in Figure \ref{depthFromShock} (recall that we are sampling from an inverse exponential distribution).

\section{Interstellar dust}
\label{app:dust}
In dusty media the increase in path length caused by scattering can allow the dust to significantly attenuate the Ly$\alpha$ flux that escapes.
Because photons of different frequencies have different scattering cross-sections the average number of scatterings they undergo will be different.
Therefore they will have different changes in path-length, and subsequently undergo different levels of dust attenuation.

\begin{figure}
\includegraphics[width=80mm]{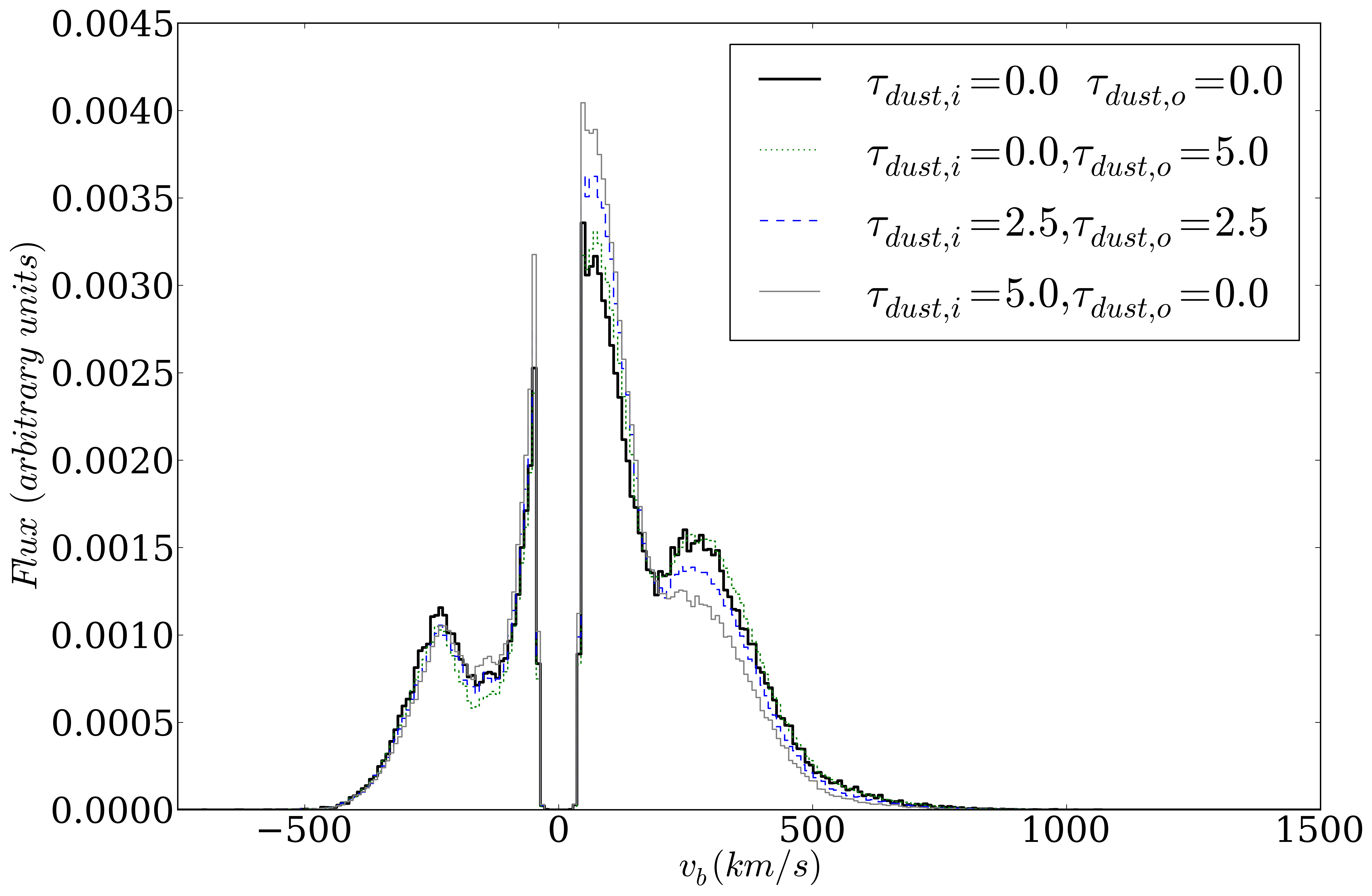}
\caption{Emergent spectrum for a shell-model with $(N_{\rm HI}^{\rm shell},\,N_{\rm HI}^{\rm stat},\,v_{\rm shell}) = (10^{19}\,{\rm cm}^{-2},\,10^{18}\,{\rm cm}^{-2},\,200{\rm km\,s}^{-1})$ including three different dust prescriptions.  For each histogram $\tau_{dust, i}$ refers to the optical depth of the inner shell, $\tau_{dust, o}$ refers to the optical depth of the outer shell.  The reference dust-free spectrum is shown with a heavy black line.}
\label{fig:B1}
\end{figure}

We take a direct approach to address this issue and perform numerical experiments where grey dust is added to either the inner, outer, or both shells.
The results are shown in Figure \ref{fig:B1}.
It is clear that the differences in the emergent spectrum caused by the different dust prescriptions are very small.

The addition of dust slightly enhances the blue bump.
This occurs because once a photon is Fermi accelerated the optical depth of the HI shells to the photon is vastly reduced and the photon easily escapes, avoiding or reducing the number of further scatterings.
Thus, a Fermi acclerated photon has a shorter total path-length so that the presence of dust affects it less than an unaccelerated photon.
This results in a slightly enhanced blue peak in the normalised spectra shown in Figure \ref{fig:B1}.

\section{Varying the HI Column Density in the Static Shell}
\begin{figure}
\includegraphics[width=80mm]{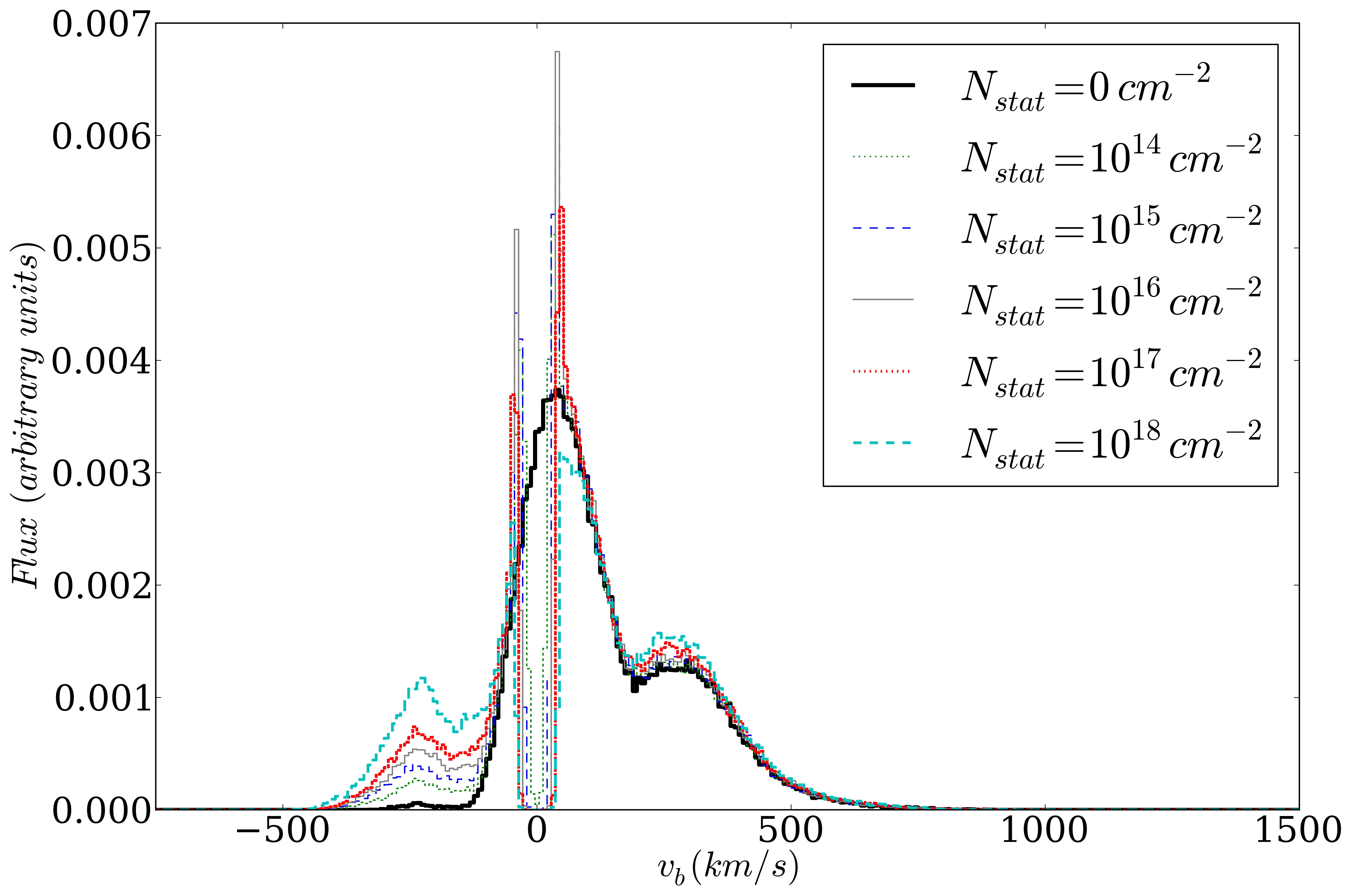}
\caption{Same as Figure~\ref{fig:B1}, but here we vary $N_{\rm HI}^{\rm stat}$ while keeping the other parameters fixed. The prominence of the blue peak reduces with HI column density, but a blue bump clearly remains even when $N_{\rm HI}^{\rm stat}=10^{15}$ cm$^{-2}$.}
\label{fig:B2}
\end{figure}

Figure~\ref{fig:B2} shows that the blue bump remains visible even if we further reduce $N_{\rm HI}^{\rm stat}$ by order(s) of magnitude.

\label{lastpage}
\end{document}